\documentclass[sigconf]{acmart}

\renewcommand\footnotetextcopyrightpermission[1]{}
\setcopyright{none}

\usepackage{tabularray}
\usepackage{xcolor}
\usepackage{balance}
\usepackage{multirow}
\usepackage[ruled,linesnumbered]{algorithm2e}
\usepackage{tcolorbox}
\usepackage[noend]{algpseudocode}
\usepackage[flushleft]{threeparttable}
\usepackage{fbox}
\usepackage{xcolor}
\usepackage{wasysym}
\usepackage{booktabs}
\usepackage{verbatim}
\usepackage{pgfplots}
\pgfplotsset{compat=1.18}
\usepackage{hyperref}
\usepackage{fontawesome}
\usepackage{tikz}
\usepackage{subfigure}
\usepackage{graphics}
\usepackage{multirow}
\usepackage{fancybox}
\usepackage{listings}
 
\AtBeginDocument{%
  \providecommand\BibTeX{{%
    \normalfont B\kern-0.5em{\scshape i\kern-0.25em b}\kern-0.8em\TeX}}}

\usepackage{wasysym}

\begin{document}

\begin{abstract}

A test oracle serves as a criterion or mechanism to assess the correspondence between software output and the anticipated behavior for a given input set.
In automated testing, black-box techniques, known for their non-intrusive nature in test oracle construction, are widely used, including notable methodologies like differential testing and metamorphic testing.
Inspired by the mathematical concept of \textit{inverse function}, we present \textit{Retromorphic Testing}, a novel black-box testing methodology.
It leverages an auxiliary program in conjunction with the program under test, which establishes a dual-program structure consisting of a forward program and a backward program.
The input data is first processed by the forward program and then its program output is reversed to its original input format using the backward program. 
In particular, the auxiliary program can operate as either the forward or backward program, leading to different testing modes.
The process concludes by examining the relationship between the initial input and the transformed output within the input domain.
For example, to test the implementation of the sine function $\sin(x)$, we can employ its inverse function, $\arcsin(x)$, and validate the equation $x = \sin(\arcsin(x)+2k\pi), \forall k \in \mathbb{Z}$.
In addition to the high-level concept of Retromorphic Testing, this paper presents its three testing modes with illustrative use cases across diverse programs, including algorithms, traditional software, and AI applications.



\end{abstract}

\title{Retromorphic Testing}
\subtitle{A New Approach to the Test Oracle Problem}


\author{Boxi Yu}
\authornote{Both authors contributed equally to this research.}
\email{boxiyu@link.cuhk.edu.cn}
\affiliation{
  \institution{School of Data Science, The Chinese University of Hong Kong, Shenzhen (CUHK-Shenzhen), China}
  \country{}
}

\author{Qiuyang Mang}
\authornotemark[1]
\email{qiuyangmang@link.cuhk.edu.cn}
\affiliation{
  \institution{School of Data Science, The Chinese University of Hong Kong, Shenzhen (CUHK-Shenzhen), China}
  \country{}
}

\author{Qingshuo Guo}
\email{qingshuoguo@link.cuhk.edu.cn}
\affiliation{
  \institution{School of Data Science, The Chinese University of Hong Kong, Shenzhen (CUHK-Shenzhen), China}
  \country{}
}

\author{Pinjia He}
\authornote{Pinjia He is the corresponding author.}
\email{hepinjia@cuhk.edu.cn}
\affiliation{
  \institution{School of Data Science, The Chinese University of Hong Kong, Shenzhen (CUHK-Shenzhen), China}
  \country{}
}

\newcommand{\etal}{{\em et al.}\xspace}
\newcommand{\ie}{{\em i.e.},\xspace}
\newcommand{\eg}{{\em e.g.},\xspace}

\newcommand{{\methodname}}{\text{Retromorphic Testing}\xspace}

\settopmatter{printacmref=false}




\keywords{Test oracle problem, black-box testing, automated testing}



\maketitle

\section{Introduction}

Test oracle plays a pivotal role in software testing~\cite{Barr2015}, serving as a critical benchmarking mechanism for assessing the correctness of software output given specific input~\cite{Howden1978}.
The significance of test oracles becomes apparent in automated validation because they establish the ultimate standard for the evaluation of system behaviors.
However, test oracles are difficult to obtain in practice due to many reasons, such as the lack of clear specifications, the complexity and dynamic nature of software, and limited resources.

To solve the test oracle problem, research in recent decades has proposed a variety of useful testing techniques~\cite{Barr2015, Pezze2014, Segura2016, Chen2018}.
Among these techniques, \textit{differential testing}~\cite{McKeeman1998} and \textit{metamorphic testing}~\cite{Chen1998} are two typical methodologies that have been widely employed because of their black-box nature and simple yet effective high-level ideas.
In differential testing, multiple systems or software versions are expected to exhibit identical behaviors for the shared functionalities.
Discrepancies in output for identical inputs among these systems indicate potential bugs in one or more systems.
Metamorphic testing generates new test inputs based on existing input-output pairs, with the output of the generated input being predictable based on the relationship between the original input and the generated one.
The detection of deviations in output relationships can highlight potential bugs in the target systems.
In addition to these black-box methodologies, Rigger and Su~\cite{rigger_intra} recently introduced \textit{intramorphic testing}, a white-box automated testing methodology, which modifies a system component and constructs a test oracle that can relate the output of the original and modified systems given the same input.

This paper introduces \textit{Retromorphic Testing}, a general black-box methodology to tackle the challenges in test oracle construction.
The core idea of Retromorphic Testing is inspired by mathematical relationships between functions and their inverse functions, expressed as $f^{-1}(f(x)) = x$.
It employs an auxiliary program to reverse the output of the software, transforming it back to its original input format.
Different from differential testing and metamorphic testing, which typically involve a system or systems with equivalent functionalities, Retromorphic Testing employs a ``dual-program structure'', comprising a forward program and a backward program, where the forward program corresponds to $f(x)$ and the backward program corresponds to $f^{-1}(x)$.
The choice of the auxiliary program, whether forward or backward, depends on the system under test.

The dual-program structure enables three \textit{testing modes}: forward mode, backward mode, and integrated mode, which is decided by the role of the system under test in the structure.
For example, to test the $\sin(x)$ function using Retromorphic Testing, the forward mode employs the test oracle $\arcsin(\sin(x)) = x, \frac{\pi}{2} \leq x \leq \frac{\pi}{2}$, where the program under test (\ie $\sin(x)$) is the forward program; while the backward mode uses the test oracle $\sin(\arcsin(x)) = x, -1 \leq x \leq 1$.
The integrated mode uses the system under test as both the forward program and backward program.
To test the $\frac{1}{x}$ function, the integrated mode can construct the test oracle $f(f(x)) = x, x \neq 0$.
The diverse testing modes provided by Retromorphic Testing bring additional flexibility to test oracle construction by assigning different roles to the programs involved. 

To demonstrate the flexibility of Retromorphic Testing, we provide six practical examples of possible implementations of Retromorphic Testing in validating algorithms, traditional software, and AI software.
We believe Retromorphic Testing has been realized in different testing scenarios by researchers or practitioners.
However, they mainly regard it as an ad hoc solution for testing specific software or as implementation details of part of the test suite. 
This paper aims to distill the high-level concept behind and provide a general definition of this methodology, which we hope will further facilitate the development and discussion of more instances of Retromorphic Testing.

In summary, this paper makes the following contributions:

\begin{itemize}
    \item It introduces \textit{Retromorphic Testing}, a black-box methodology to tackle the test oracle problem.

    \item It discusses three \textit{testing modes} enabled by the dual-program structure in Retromorphic Testing.

    \item It provides six instances of Retromorphic Testing.



\end{itemize}

\section{Preliminaries}

\begin{figure*}[ht]
    \centering
    \includegraphics[width=0.95\linewidth]{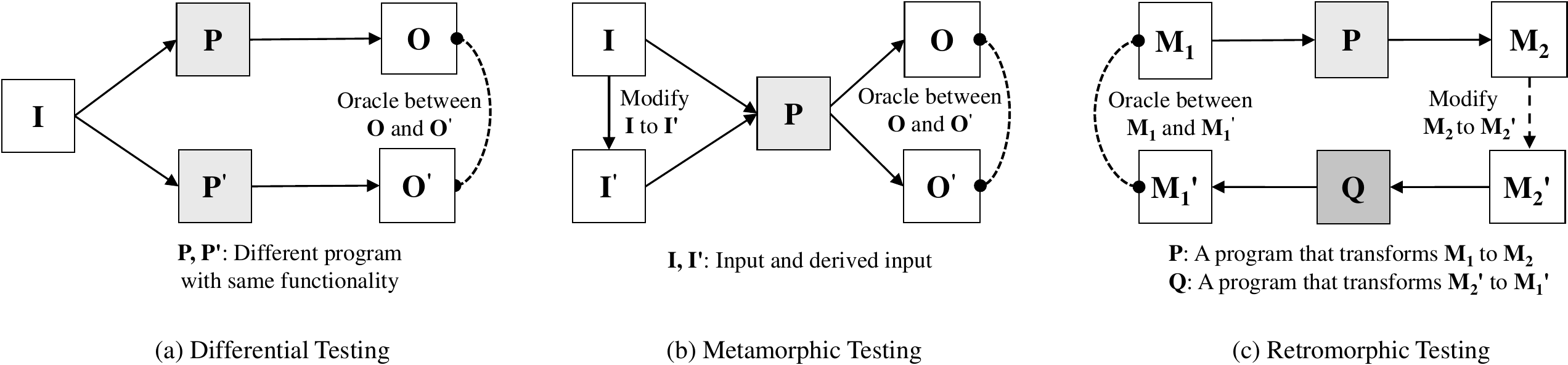}
    \caption{Differential testing, metamorphic testing, and Retromorphic Testing in comparison.}
    \label{fig:threeTest}
\end{figure*}

\subsection{Test Oracles}
The test oracle problem stands as one of the most significant challenges in the realm of software testing. A test oracle serves as a mechanism designed to verify the correctness of a system's output for a set of specific inputs~\cite {Howden1978}.
This problem encompasses the complex task of automating the differentiation between expected and potentially erroneous system behavior, 
which poses a significant bottleneck in the development of more comprehensive testing methodologies.
Ideally, a system would be accompanied by a detailed specification of its expected behaviors, or its code would contain predefined conditions, allowing for an automated test oracle to verify outputs without human intervention.
However, in many real-world scenarios, software developers lack such luxuries, leading to manual verification of system behaviors by handmade test cases, which can be extremely time-consuming and inefficient in bug detection.
To mitigate this challenge, several approaches that generate ``\textit{partial test oracles}''~\cite{Barr2015} have been proposed, which can still automatically validate system output from a part of inputs.
The most influential ones are differential testing~\cite{McKeeman1998} and metamorphic Testing~\cite{Chen1998}.

\subsection{Terminology}
For existing testing techniques such as differential testing~\cite{McKeeman1998} and metamorphic testing~\cite{Chen1998}, test oracles were defined to validate outputs for a set of inputs~\cite{Howden1978}.
In this paper, we continue to use this terminology in these approaches, where the target program is denoted as $P$, the input as $I$, and the output as $O$, where $O = P(I)$.

However, we will not use the original terminology in Retromorphic Testing due to its dual-program structure, but consider the program execution as a transition between two modalities.
In this scenario, we use $M_i, i \in \{1, 2\}$ to represent the data belonging to the $i$-th modality.
We define the forward program $P$ as the program that transforms the data from $M_1$ to $M_2$ ($M_2 = P (M_1)$), and the backward program $Q$ that transforms the data from $M_2$ back to $M_1$ ($M_1 = Q(M_2)$).

\subsection{Black-box Testing Techniques}
Among existing test oracle approaches, black-box testing techniques are a set of methodologies where the tester is not required to know the internal knowledge or implementation details of the target system.
Instead, the focus is on evaluating system behaviors based on its inputs and outputs.
Among the prominent black-box testing techniques are differential testing~\cite{McKeeman1998} and metamorphic testing~\cite{Chen1998}.
Differential testing involves comparing the outputs of two or more versions of a software application to identify discrepancies, often used to detect inconsistencies in versions of compilers or other tools that should produce equivalent results.
With its effectiveness, differential testing has been applied to a variety of domains, such as Rust compilers~\cite{sharma2023RustTest}, model counters~\cite{Usman2021TestMC}, Java Virtual Machines (JVMs)~\cite{Chen2016, Chen2019}, database engines~\cite{Slutz1998, grand}, and deep-learning libraries~\cite{yang2023faddll, deng2023large}.
Metamorphic testing is a technique where the tester identifies certain properties that should remain invariant under specific transformations of the input.
If these properties change after the transformation, a defect might be signaled.
This approach is particularly useful in scenarios where sufficient test oracles not be available or are too expensive to use.
Metamorphic testing is also used for testing various systems such as compilers~\cite{Le2014, Li2023SOSP}, database engines~\cite{Rigger2020, Rigger2020b, kamm2023TGDB}, SMT solvers~\cite{Winterer2020b}, Android apps~\cite{Su2021, sun2021}, quantum computing platforms~\cite{paltenghi2023MTQuant, abreu2022}, and AI systems~\cite{ Wang2023ValidatingMC, sun2020TransRepair, yu2023IC, yu2023NER, Shuai2020}.

In Fig.~\ref{fig:threeTest} (a) and (b), we give the workflow of differential testing and metamorphic testing, respectively.
Differential testing uses a fixed input $I$ for programs $P$ and $P'$ of the same functionality and validates whether their outputs, \textit{i.e.,} $O = P (I)$ and $O' = P'(I)$, are equal.
When there is a discrepancy between $O$ and $O'$, it would indicate bugs in either $P$ or $P'$.
Unlike differential testing which uses the same input for different programs, metamorphic testing mutates the input $I$ to $I'$ and feeds them into one given program.
It is based on the idea of defining \textit{metamorphic relations (MRs)} that capture the expected relationships between the outputs $O = P(I)$ and $O' = P(I')$ of the program.


\section{Retromorphic Testing}

\subsection{Testing Principle}


As illustrated in Fig.~\ref{fig:threeTest} (c), Retromorphic Testing adopts a \textit{dual-program structure} consisting of a forward program $P$ and a backward program $Q$.
Its testing process starts with $M_1$ (\textit{e.g.,} trigonometric values), which serves as the generated input for the forward program $P$ (\textit{e.g.,} $y = \arcsin(x)$).
Using $P$, Retromorphic Testing transforms $M_1$ into output $M_2$ (\textit{e.g.,} arcs). 
Following this, a mutation method denoted as $Mutate$ is applied to convert $M_2$ into $M_2'$, where $M_2' = Mutate(M_2)$ (\textit{e.g.,} $\theta' = \theta + 2k\pi, k \in \mathbb{Z}$).
Notably, the function $Mutate$ could be an identity function, making $M_2$ and $M_2'$ identical.
The mutated data, $M_2'$, is then input into program $Q$ (\textit{e.g.,} $y = \sin(x)$), resulting in $M_1'$. 
The Retromorphic Relation exists between $M_1$ and $M_1'$ (\textit{e.g.,} $x = \sin(\arcsin(x)+2k\pi), k \in \mathbb{Z}$).
If this Retromorphic Relation does not hold, a bug is detected in program $P$ or $Q$. 

\subsection{Testing Modes}
In the dual-program structure of Retromorphic Testing, the target program can serve as either the forward program $P$ or the backward program $Q$.
This results in three testing modes for Retromorphic Testing, namely \textit{forward mode}, \textit{backward mode}, and \textit{integrated mode}. 
Each mode has a distinct process for input generation and result validation, offering developers flexibility in designing generators and validating programs.

\paragraph{\textbf{Forward Mode.}}
In the forward mode of Retromorphic Testing, the target program serves as the forward program $P$. 
Here, our task is to find a backward program $Q$, which transforms the output produced by $P$ back into the same format as $P$'s input.
For instance, if we want to test a program like $P \colon y = \sin(x)$ (assuming it contains bugs) using the forward mode, we can select a backward program like $Q \colon y = \arcsin(x)$ (assuming it is bug-free). 
We generate a set of arc inputs within the range of $-\frac{\pi}{2}$ to $\frac{\pi}{2}$ and pass them through $P$ to obtain the trigonometric values. 
After that, we send these values to $Q$, transforming them back to the arcs (\textit{i.e.,} $\arcsin(\sin(x)) = x$).
By comparing the generated arcs with the arcs returned from $Q$, we can identify potential bugs in the target program.
In the forward mode, we need to generate the data and validate the results, which are both in the input modality of the target program (\textit{e.g.}, arcs in the above example).



\paragraph{\textbf{Backward Mode.}}
In the back mode of Retromorphic Testing, the target program serves as the backward program $Q$. 
Here, our task is to find a forward program $P$, where the target program can transform the output produced by $P$ back into the same format as $P$'s input.
Similar to the forward mode, if we want to test a program like $Q \colon y = \sin(x)$ (assuming it contains bugs) using the backward mode, we can select a back program like $P \colon y = \arcsin(x)$ (assuming it is bug-free). 
We generate a set of trigonometric value inputs within the range of $-1$ to $1$ and pass them through $P$ to obtain the arcs. 
We can also mutate these arcs by adding $2k\pi$ ($k \in \mathbb{Z}$) to them.
After that, we send these mutated arcs to $Q$, transforming them back to the trigonometric values (\textit{i.e.,} $\sin(\arcsin(x)+2k\pi) = x$).
By comparing the generated trigonometric values with the values returned from $Q$, we can identify potential bugs in the target program.
In the backward mode, we need to generate the data and validate the results, which are both in the output modality of the target program (\textit{e.g.}, trigonometric values in the above example).

\paragraph{\textbf{Integrated Mode.}}
In the integrated mode of Retromorphic Testing, the target program serves as both the forward program $P$ and the backward program $Q$ simultaneously (if possible), which means that we do not need to find another auxiliary program in this mode.
In particular, $P$ and $Q$ could be the same system or two different components of the same system.
For example, if we want to test a program like $P \colon y = \frac{1}{x}$, we can simply select itself as the other program (\textit{i.e.,} $Q \colon y = \frac{1}{x}$).
We generate a set of non-zero number inputs and pass them through $P$ and $Q$ sequentially (\textit{i.e.,} $x = \frac{1}{1/x}$).
By comparing the generated number with the number returned from $Q$, we can identify potential bugs in the target program.
In the integrated mode, the process of the data generation and results validation can be in either the input or output modality of the target program (\textit{e.g.,} non-zero values in the above example).

\subsection{Examples of Different Testing Methodologies}\label{sec:motiva}

To outline the existing techniques and our idea, let us assume a typical use case, namely that we want to test the implementation of the \textit{Discrete Fourier Transform (DFT)}. 
DFT is a mathematical technique that has been widely used in practice, such as signal processing, image compression, and audio processing.
DFT transforms a sequence from the time domain (or spatial domain) to the frequency domain.
Given a time-domain sequence of $N$ numbers $x_0, x_1, \ldots, x_{N-1}$, the sequence after DFT is a frequency-domain sequence of $N$ complex numbers given by:

\begin{equation}
X_k = \sum_{n=0}^{N-1} x_n \cdot \exp(-j \frac{2\pi}{N} k n),
\end{equation}\label{equa:dft}
for $k$ = $0$, $1$, \ldots, $N-1$, where:
\begin{itemize}
    \item \( X_k \) are the frequency components of the sequence.
    \item \( \exp \) is the base of the natural logarithm.
    \item \( j \) is the imaginary unit.
\end{itemize}

The inverse operation, which transforms the frequency-domain sequence back to the time domain, is called the \textit{Inverse Discrete Fourier Transform (IDFT)}.
IDFT is defined as:
\begin{equation}
x_n = \frac{1}{N} \sum_{k=0}^{N-1} X_k \cdot \exp(j \frac{2\pi}{N} k n)
\end{equation}\label{euqa:idft}

Let us further assume that we made a mistake when implementing the \texttt{discrete\_fourier\_transform()} function, as illustrated in Listing \ref{lst:dft};
The coefficient is incorrectly typed as $-1j$ rather than $-2j$.  
In the subsequent paragraphs, we discuss how the instantiations of existing techniques and an instantiation of our proposed Retromorphic Testing, could find the bug.
In practice, we expect that Retromorphic Testing will be realized that can find bugs that are overlooked, or difficult to find, by other testing approaches.

\begin{figure}
\begin{lstlisting}[caption={A Python implementation of Discrete Fourier Transform }, label=lst:dft, belowcaptionskip=\baselineskip, belowskip=-2.5ex]
import numpy as np

def discrete_fourier_transform(sequence, opt = 1):
    N = len (sequence)
    sequence_new = [0 for _ in range ( N )]
    for k in range (N):
        for n in range (N):
            sequence_new [k] += sequence [n] * np.exp (opt * |\textbf{\textcolor{red}{-1j}}| |\textcolor{red}{\faBug}| * np.pi / N * k * n) # Bug: should be -2j
    if(opt == -1):
        for i in range(len(sequence_new)):
            sequence_new[i] = sequence_new[i] / N
    return sequence_new
\end{lstlisting}
\end{figure}\label{fig:py_impl_dft-idft}


\paragraph{\textbf{Manual Testing.}}

For manual testing, we need to generate the test cases and their corresponding expected outputs.
Listing \ref{lst:reg} shows a test case in which we use manual testing to detect the bug of the DFT algorithm.
The algorithm under test transforms a sequence [1, 0, 1, 0] into [2, 1, 0, 0], whereas the expected output after DFT is [2, 0, 2, 0], thus the bug of the algorithm is uncovered.
While manual testing is effective and widely utilized, it demands considerable effort from developers to create test cases and expected outputs.

\begin{figure}
\begin{lstlisting}[caption={A manually-written test case for Discrete Fourier
Transform.}, label=lst:reg, belowcaptionskip=\baselineskip, belowskip=-2.5ex]
time_sequence = [2 , 0 , 1 , 0]
frequency_sequence = discrete_fourier_transform(time_sequence)
frequency_sequence = [int(_.real) for _ in frequency_sequence]
assert frequency_sequence == [2 , 0 , 2 , 0]
\end{lstlisting}
\end{figure}

\paragraph{\textbf{Differential Testing.}}
Differential testing ~\cite{McKeeman1998} validates a set of systems that implement the same semantics, by comparing their output for a given input.
Given that the test oracle requires no human in the loop, it can be effectively paired with automated test generation.
For example, in Listing~\ref{lst:dif}, we generate random sequences as test input.
It is worth noticing that we use an infinite loop in this example; in the practice of using differential testing, it would be reasonable to set a timeout or run the tests for a fixed number of iterations.
For an input sequence like [1, 0, 1, 0], differential testing reveals a discrepancy between the output of the two Fourier transform algorithms, revealing the bug.

\begin{figure}
\begin{lstlisting}[caption={Differential testing using multiple implementations of transforming algorithms.}, label=lst:dif, belowcaptionskip=\baselineskip, belowskip=-2.5ex]
eps = 1e-10
transform_algorithms = [discrete_fourier_transform, fast_fourier_transform]
while True :
    time_sequence = get_random_sequence()
    frequency_sequences = [alg(time_sequence.copy()) for alg in transform_algorithms]
    frequency_sequences = [[_.real for _ in frequency_sequence] for frequency_sequence in frequency_sequences]
    all_same = all(all(abs(frequency_sequence[i] - frequency_sequences[0][i]) < eps for i in range(len(frequency_sequence))) for frequency_sequence in frequency_sequences)
    assert all_same
\end{lstlisting}
\end{figure}

\paragraph{\textbf{Metamorphic Testing.}}


To adopt metamorphic testing ~\cite{Chen1998} for testing DFT, we can deduce an MR from the definition formula of DFT.
Assume $x_0, x_1, \ldots, x_{N-1}$ is the input time-domain sequence, and $X_0, X_1, \ldots, X_{N-1}$ is the correct output frequency-domain sequence after DFT.
When a constant $c$ is added to the first element (\textit{i.e.}, $x_0$) to generate a new input sequence $x_0+c, x_1, \ldots, x_{N-1}$, the output frequency-domain sequence should be $X_0+c, X_1+c, \ldots, X_{N-1}+c$.
We can use the property of DFT as an MR to verify the implementation of DFT.

Using the implementation in Listing~\ref{lst:dft}, when the input sequence is [1, 0, 1, 0], the output sequence is [2, 1, 0, 0].
When passing [2, 1, 0, 1] as the input by adding a constant 1 at the first element, the output is [3, 2, 1, 1], which matches the expected output sequence based on the original output [2, 1, 0, 0].
Therefore, metamorphic testing with this MR failed to identify this bug.

\begin{figure}
\begin{lstlisting}[caption={Metamorphic testing by comparing whether the
relative difference is maintained for a modified polynormial}, label=lst:met, belowcaptionskip=\baselineskip, belowskip=-2.5ex]
eps = 1e-10
while True :
    time_sequence = get_random_sequence() 
    frequency_sequence = discrete_fourier_transform(time_sequence)
    c = random.random()
    time_sequence[0] += c
    frequency_sequence_new = discrete_fourier_transform(time_sequence)
    all_same = all(abs(frequency_sequence_new[i].real - frequency_sequence[i].real - c) < eps for i in range(len(frequency_sequence))) 
    assert all_same
\end{lstlisting}
\end{figure}

\paragraph{\textbf{Retromorphic Testing}}


Illustrated in Listing~\ref{lst:ret}, Retromorphic Testing first utilizes DFT to convert the randomly generated time-domain input sequence, $x_0, x_1, \ldots, x_{N-1}$, into a frequency-domain sequence $X_0, X_1, \ldots, X_{N-1}$.
Given the inherent properties of DFT and IDFT, it is anticipated that the two time-domain sequences (\textit{i.e.,} $x_0, x_1, \ldots, x_{N-1}$ and $x_0', x_1', \ldots, x_{N-1}'$) hold the following relationship:
\begin{equation}
\begin{aligned}
x_i' &= x_i, \text{for all } 0 \leq i < N.
\end{aligned}
\end{equation}\label{equa:idftRR}

This interrelation can serve as the Retromorphic Relation that validates the DFT implementation.
For example, consider a time-domain sequence $x$ = [1, 0, 1, 0].
Employing DFT, we expect to acquire the corresponding frequency-domain sequence $X$, which should ideally be [2, 0, 2, 0].
However, the implementation shown in Listing~\ref{lst:dft} fails to yield the correct frequency-domain sequence. 
Instead, it returns [2, 1, 0, 0].
The realization in Listing~\ref{lst:ret} of Retromorphic Testing can easily detect this bug.
When feeding the frequency-domain sequence into the IDFT program, it yields a time-domain sequence $x' = [0.75, 0.68, 0.5, 0.32]$.
However, $x'$ is expected to be [1, 0, 1, 0] according to the Retromorphic Relation defined in Equation~\ref{equa:idftRR}, thereby we have detected the bugs in the DFT realization.

In addition, we can modify the frequency-domain sequence from the DFT program to enhance the test oracle, similar to metamorphic testing.
To achieve this, we add a constant $c$ to each element of the frequency-domain sequence to get the modified sequence $X_0+c, X_1+c, \ldots, X_{N-1}+c$, and employ the IDFT to revert the modified sequence $X_0+c, X_1+c, \ldots, X_{N-1}+c$ back to the time domain, yielding the sequence $x_0' + c, x_1', \ldots, x_{N-1}'$.
Therefore, the Retromorphic Relation will be changed as:

\begin{equation}
\begin{aligned}
x_0' &= x_0 + c , \\
x_i' &= x_i, \text{for all } 1 \leq i < N.
\end{aligned}
\end{equation}\label{equa:idftRRR}

For the above example, when we add constant 1 to each element of the erroneous frequency-domain sequence [2, 1, 0, 0] to get the modified sequence [3, 2, 1, 1] and feed the modified sequence into the IDFT program, it yields a time-domain sequence $x' = [1.75, 0.93, 0.5, 0.57]$.
$x'$ is expected to be [2, 0, 1, 0] according to the Retromorphic Relation defined in Equation~\ref{equa:idftRRR}, thereby we have also detected the bugs in the DFT realization.


\begin{figure}
\begin{lstlisting}[caption={Retromophic testing by comparing whether the
relative difference is maintained for the inversed transformation}, label=lst:ret, belowcaptionskip=\baselineskip, belowskip=-2.5ex]
eps = 1e-10
while True:
    time_sequence = get_random_sequence()
    frequency_sequence = discrete_fourier_transform(time_sequence)
    
    #Without Modification
    time_sequence_new = discrete_fourier_transform(frequency_sequence, -1)
    all_same = all(abs(time_sequence_new[i].real - time_sequence[i].real) < eps for i in range(len(time_sequence))) 
    assert all_same
    
    #With Modification
    c = random.random()
    for i in range(len(frequency_sequence)):
        frequency_sequence[i] += c
    time_sequence_new = discrete_fourier_transform(frequency_sequence, -1)
    all_same = all(abs(time_sequence_new[i].real - time_sequence[i].real) < eps for i in range(1, len(time_sequence)) and abs(time_sequence_new[0].real - time_sequence[0].real) < eps) 
    assert all_same
\end{lstlisting}
\end{figure}

\section{Testing Scopes}\label{sec:test_scope}

\begin{table*}
\centering
\caption{Overview of six testing scopes using Retromorphic Testing}
\label{tab:overview}
\resizebox{0.999\linewidth}{!}{
\begin{tblr}{
  row{even} = {c},
  row{1} = {c},
  row{3} = {c},
  row{7} = {c},
  cell{5}{1} = {c},
  cell{5}{3} = {c},
  cell{5}{4} = {c},
  cell{5}{5} = {c},
  vline{2,4} = {-}{},
  hline{1-2,8} = {-}{},
}
\textbf{Testing Scope}                             & \textbf{Software Types} & \textbf{Testing Mode} & \textbf{Forward Program}   & \textbf{Backward Program}  \\
\textbf{Integer factorization}                     & Algorithm              & Forward               & \textbf{Pollard's rho*}             & Integer multiplication     \\
\textbf{Infix, prefix, and postfix transformation} & Algorithm             & Integrated            & \textbf{Expression transformation (Prefix$\rightarrow$Postfix)*} & \textbf{Expression transformation (Postfix$\rightarrow$Prefix)*} \\
\textbf{Database Management System}                & Conventional software   & Backward              & Query generator            & \textbf{Database system*}           \\
\textbf{Java decompiler}                           & Conventional software   & Backward              & Java compiler              & \textbf{Java decompiler*}           \\
\textbf{Machine translation}                       & AI~application          & Integrated            & \textbf{Google Translate (EN$\rightarrow$ZH-CN)*}          & \textbf{Google Translate (ZH-CN$\rightarrow$EN)*}          \\
\textbf{Image captioning and generation}           & AI~application          & Backward              & DALL-E2                    & \textbf{Azure's image captioning*}  
\end{tblr}
}
\begin{tablenotes}
      \footnotesize
      \item `\textbf{*}' indicates the target system under test.
\end{tablenotes}
\end{table*}\label{tab:oview_test_scope}

In this section, we present six instances of Retromorphic Testing techniques across three testing scopes, with two examples provided for each testing scope.
Table~\ref{tab:oview_test_scope} shows the statistics.
We give one example of forward mode, three examples of backward mode, and two examples of integrated mode.


\paragraph{Example 1: Integer factorization}

Integer factorization is the process of finding the prime numbers that multiply together to give a certain integer. 
This is a fundamental concept in number theory and has various applications in both mathematics and computer science such as cryptography, primality testing, cryptanalysis, digital signatures, codebreaking, and computational complexity ~\cite{int_cry,int_tes,int_ppp}. 
The most straightforward way to factorize an integer is to test all the prime numbers not greater than the square root of it, which requires a time complexity of $O(\sqrt{N})$, where $N$ is the input integer. 
However, the integer that needs to be factorized is usually extremely large in practice, making this algorithm unacceptable. 
To effectively factorize large integers, programs for practical applications are much more complicated than straightforward algorithms. 

Pollard's rho is one of the effective algorithms for integer factorization.
As shown in Listing \ref{lst:ifa}'s Python implementation, it employs a divide-and-conquer approach by iteratively selecting a factor of $N$, denoted as $d$.
Then, the procedure is divided into two sub-processes for $\frac{N}{d}$ and $d$ respectively, continuing this division until it identifies a prime number.
However, in this program, the calculation of the selected factor $d$ is erroneously implemented as \texttt{gcd(abs(x - y), \textcolor{red}{x})}, instead of the correct version \texttt{gcd(abs(x - y), n)}.

To test this program, we can utilize Retromorphic Testing based on the fact that prime numbers multiply together to the given integer.
Assume $N$ is the input integer to be factorized, and $p_0, p_1, \ldots, p_{k}$ are the correct output prime numbers after factorization. 
The test oracle can be formulated as $N = \prod \limits_{i = 0}^{k} p_{i}$, which is demonstrated in Listing~\ref{lst:rif}.
For example, when taking $12$ as the input integer, the correct factorization should be $[2, 2, 3]$. 
However, the implementation in Listing \ref{lst:ifa} returns $[2, 2, 2]$.
When multiply these numbers together, we get the result of $8$, rather than the original input $12$, breaking the Retromorphic Relation. 
 
\begin{figure}
\begin{lstlisting}[caption={Pollard's rho algorithm for integer factorization}, label=lst:ifa, belowcaptionskip=\baselineskip, belowskip=-2.5ex]
import random

def gcd(a, b):
    while b:
        a, b = b, a % b
    return a

def pollards_rho(n):
    if n == 1:
        return []
    if n % 2 == 0:
        return [2] + pollards_rho(n // 2)
    
    def rho(x, c):
        return (x * x + c) % n
    
    x = random.randint(1, n - 1)
    y = x
    c = random.randint(1, n - 1)
    d = 1
    while d <= 1:
        x = rho(x, c)
        y = rho(rho(y, c), c)
        d = gcd(abs(x - y), |\textbf{\textcolor{red}{x}}|) |\textcolor{red}{\faBug}|  # Bug: should be gcd(abs(x - y), n)
    if d == n:
        return [n]
    else:
        return pollards_rho(d) + pollards_rho(n // d)
\end{lstlisting}
\end{figure}
\begin{figure}
\begin{lstlisting}[caption={Retromorphic Testing for integer factorization}, label=lst:rif, belowcaptionskip=\baselineskip, belowskip=-2.5ex]
while(True):
    N = get_random_integer()
    factors = pollards_rho(N)
    result = 1
    for prime in factors:
        result *= prime
    assert(result == N)
\end{lstlisting}
\end{figure}

\begin{figure}[!h]
\begin{lstlisting}[caption={Code for transforming between prefix and postfix notation}, label=lst:ast, belowcaptionskip=\baselineskip, belowskip=-2.5ex]
def is_operator(char):
    return char in "+-*/"

def postfix_to_prefix(postfix_expression):
    stack = []
    
    for char in postfix_expression:
        if char.isalnum():  # Operand
            stack.append(char)
        elif is_operator(char):
            operand|\textbf{\textcolor{red}{1}}| |\textcolor{red}{\faBug}| = stack.pop()
            operand|\textbf{\textcolor{red}{2}}| |\textcolor{red}{\faBug}| = stack.pop() #swap operand1 and operand2 for correct version
            prefix = f"{char}{operand1}{operand2}"
            stack.append(prefix)
    
    return stack[0]

def prefix_to_postfix(prefix_expression):
    stack = []

    for char in reversed(prefix_expression):
        if char.isalnum():  #Operand
            stack.append(char)
        elif is_operator(char):
            operand1 = stack.pop()
            operand2 = stack.pop()
            postfix = f"{operand1}{operand2}{char}"
            stack.append(postfix)

    return stack[0]
\end{lstlisting}
\end{figure}

\begin{figure}
\begin{lstlisting}[caption={Retromophic testing for transforming between prefix and postfix notation}, label=lst:astr, belowcaptionskip=\baselineskip, belowskip=-2.5ex]
postfix_expression = get_random_postifx()

assert(postfix_expression == prefix_to_postfix(postfix_to_prefix(postfix_expression)))

\end{lstlisting}
\end{figure}


\paragraph{Example 2: Infix, prefix, and postfix transformation}

The Abstract Syntax Tree (AST) is a tree-like data structure employed to represent the syntactic information of source code within programming languages. 
For example, when considering the arithmetic expression \texttt{5 + 6 * a}, the corresponding AST can be denoted by the construct \texttt{Operator(`+', constant(5), Operator(`*', constant(6), a))}.
The process of rendering an AST into a comprehensible format frequently involves adopting infix, prefix, or postfix notations. 
These notations dictate the positioning of operators concerning their operands in binary operations, whether they appear between, before, or after the operands. 
For example, \texttt{5 + 6 * a} is the infix notation of the arithmetic expression, \texttt{+ 5 * 6 a} and \texttt{5 6 a * +} are the corresponding prefix and postfix notation. 

Transformations between different notations are used by different tasks in practice.
Listing \ref{lst:ast} shows a Python implementation of transformation between postfix notation and prefix notation as \texttt{postfix\_to\_prefix} and \texttt{prefix\_to\_postfix}. 
In its implementation of \texttt{postfix\_to\_prefix}, the positions of \texttt{operand1} and \texttt{operand2} are erroneously exchanged.

Similar to Integer factorization, test oracles for this program can also be derived from Retromorphic Testing. 
As shown in Listing~\ref{lst:astr}, we can verify it based on the principle that if we convert a given postfix notation expression to its equivalent prefix notation and then convert the result back to postfix notation, it should match the original expression.
The test oracle can be formulated as Equation~\ref{equ:ast}, where $S$ denotes a postfix notation.
\begin{equation}
\label{equ:ast}
    S = \texttt{prefix\_to\_postfix(postfix\_to\_prefix(} S \texttt{))}
\end{equation}

For example, if we input the postfix notation \texttt{5 6 a * +} into \texttt{postfix\_to\_prefix}, it will return a prefix notation \texttt{+ * a 6 5}.
When converting this result back through \texttt{prefix\_to\_postfix}, the returned postfix notation is \texttt{a 6 * 5 +}, which is different from the original expression \texttt{5 6 a * +}, breaking the Retromorphic Relation.

\paragraph{Example 3:  Database Management System}

\begin{figure}[t]
    \centering
    \includegraphics[width=1\linewidth]{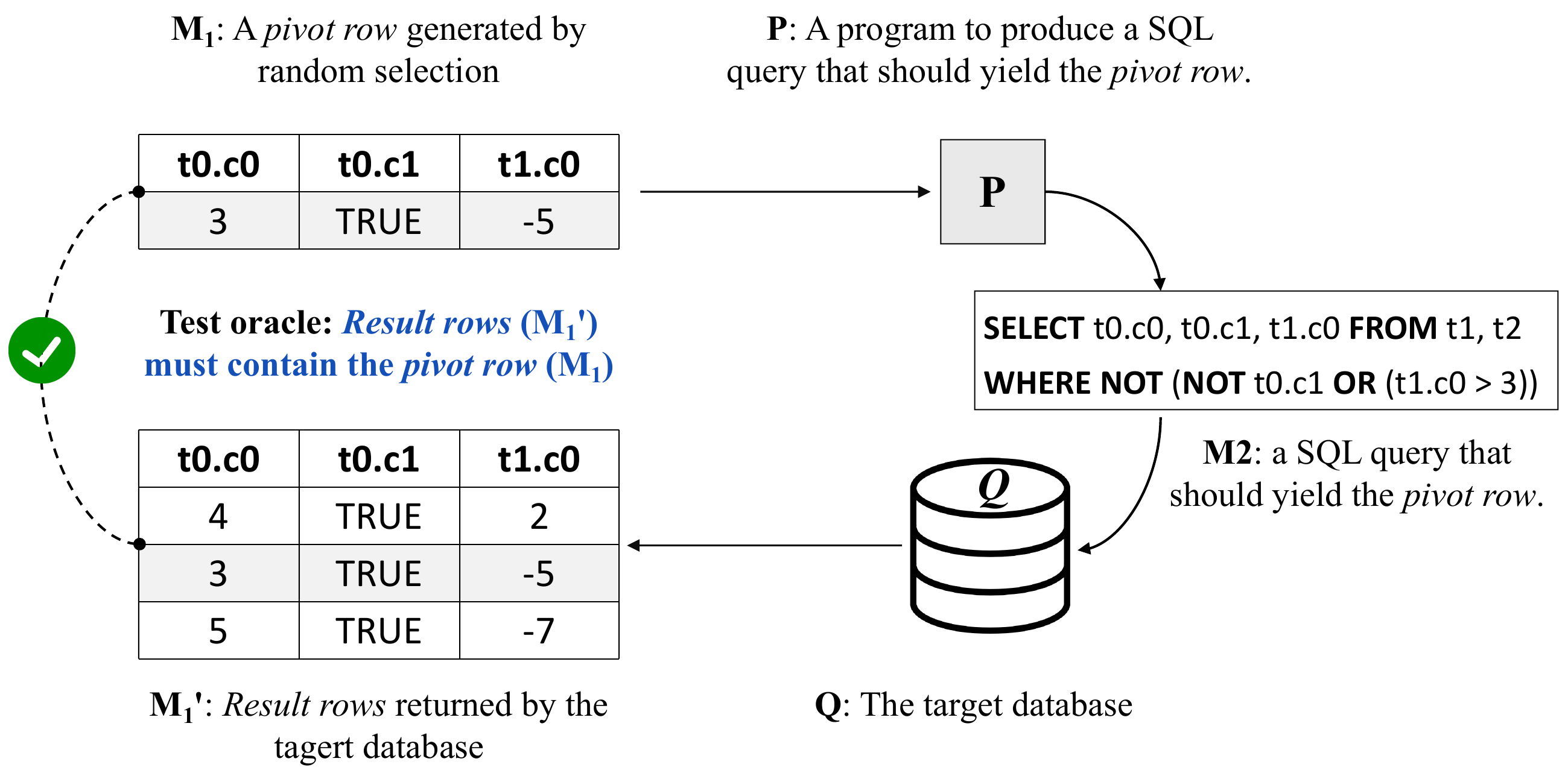}
    \caption{An illustrative example about using Retromorphic Testing to test database systems}
    \label{fig:PQS}
\end{figure}

In the context of Database Management Systems (DBMS), Rigger and Su introduced the concept of \textit{Pivoted Query Synthesis (PQS)}~\cite{rigger2020testing} as an effective testing approach to uncover logic bugs in these systems. Although the original paper describes PQS as a testing technique for a specific system, we think it can be regarded as an instance of the general Retromorphic Testing methodology. 
We also discussed with one author of PQS and he agrees that PQS should be categorized as a Retromorphic Testing technique.

As depicted in Fig.\ref{fig:PQS}, the target database serves as the backward program $Q$ in Retromorphic Testing. 
The PQS testing approach initiates with the generation of a \textit{pivot row} (\textit{e.g.,} \text{[}t0.c0: 3, t0.c1: TRUE, t1.c0: -5] in Fig.\ref{fig:PQS}), which is in the modality of rows. 
Subsequently, a simple program $P$ based on AST will be utilized to generate an SQL query based on the pivot row, ensuring that the data of the pivot row is retrieved by the query. 
This query exists in the modality of SQL queries and is then passed to the target database, reverting the modality back into rows.
The underlying Retromorphic Relation here asserts that the result obtained from the database should encompass the pivot row.


\begin{figure}[t]
    \centering
    \includegraphics[width=1\linewidth]{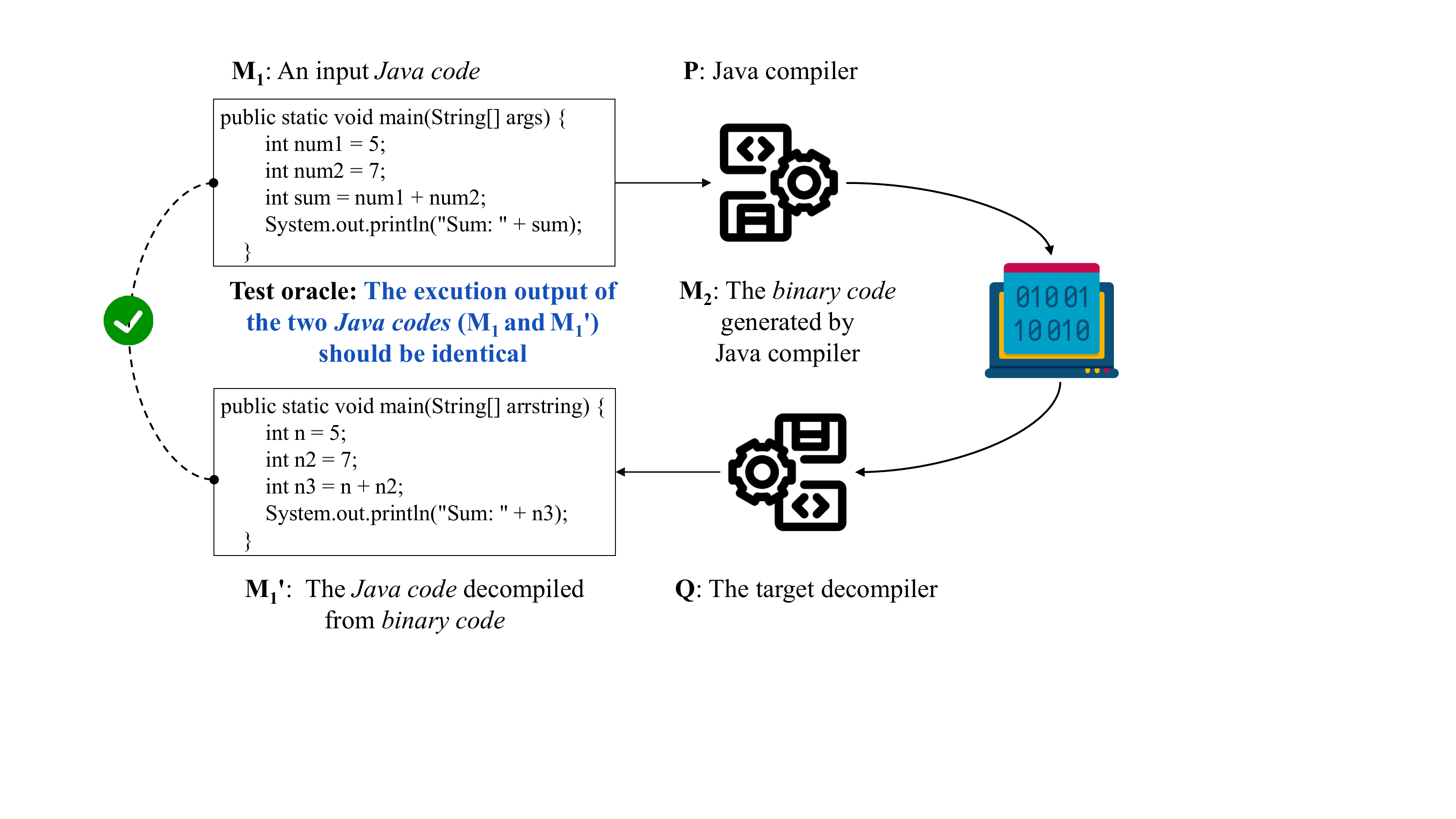}
    \caption{An illustrative example about using Retromorphic Testing to test decompiler.}
    \label{fig:decompiler}
\end{figure}

\paragraph{Example 4: Java decompiler}
While the Java compiler is renowned for its reliability, Java decompilers occasionally exhibit susceptibility to errors or inaccuracies during the transformation of compiled bytecode back into human-readable source code. These discrepancies may arise due to the intricate nature of bytecode optimization, the diversity of Java language constructs, and the inherent complexity of decompilation.
There is an example where we adopt Retromorphic Testing for testing the Java decompiler, as shown in Fig~\ref{fig:decompiler}.
As an illustration, we use $Comp$ to denote the Java compiler, and $DeComp$ to denote the Java decompiler.
In this example, we use javac 17.0.8 as the Java compiler and CFR 0.150 as the Java decompiler.\footnote{https://www.benf.org/other/cfr/}
We compile the input Java code $M_1$ into 
binary code $M_2$ and decompile $M_2$ back to Java code $M_1'$.
The test oracle between $M_1$ and $M_1'$ is their execution output.
In the example, both Java code $M_1$ and $M_1'$ outputs ``\texttt{Sum:12}'', thus the decompiler works properly in this specific test case.

\paragraph{Example 5:  Machine translation}

\begin{figure}[t]
    \centering
    \includegraphics[width=1\linewidth]{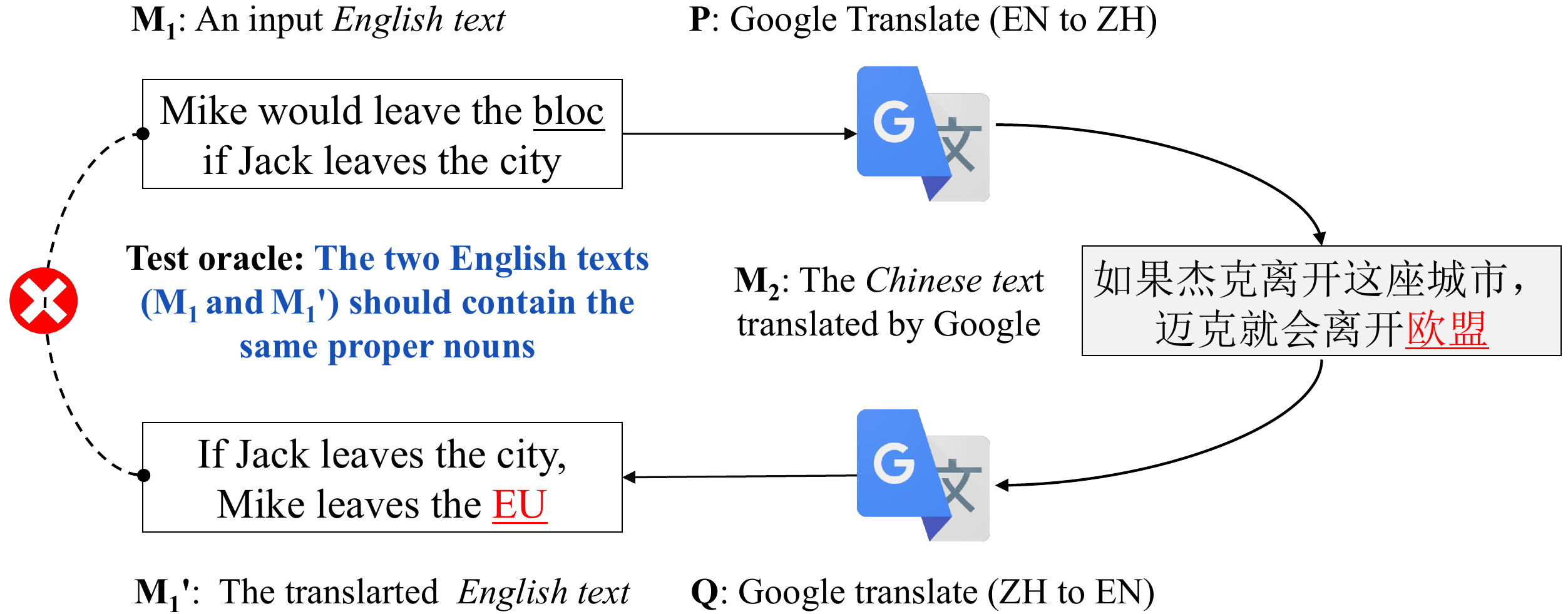}
    \caption{An illustrative example about using Retromorphic Testing to test Google Translation.}
    \label{fig:trans}
\end{figure}

Machine translation software aims to fully automate the process of translating text from a source language into a target language. 
Machine translation software often provides support for multiple languages.
Fig.~\ref{fig:trans} shows using Retromorphic Testing for testing Google Translate, where $P$ is defined as the program that translates English to Chinese, and $Q$ is the program that translates Chinese to English.
Intuitively, if we translate an English sentence $M_1$ into a Chinese sentence $M_2$ and then translate it back to an English sentence $M_1'$, the meaning of $M_1$ and $M_1'$ should remain unchanged.
We adopt a Retromorphic Relation that the proper nouns of the two sentences should be identical, which is defined below:

\begin{equation}
    PN(M_1) = PN(M_1'),
\end{equation}

where $PN(M)$ represents the proper nouns of the sentence $M$.
In the realization of this case, we adopt NLTK~\cite{bird-loper-2004-nltk}, a widely adopted NLP toolkit for extracting the proper nouns of the sentence, and get the results that $PN(M_1)$ is $\{$\textit{Mike}, \textit{Jack}$\}$, while $PN(M_1')$ is $\{$\textit{Mike}, \textit{Jack}, \textit{EU}$\}$.
Therefore, the Retromorphic Relation is not followed and there is a translating issue in Google Translate.
A similar approach has been explored by Gao \etal \cite{BTtestMT2023}.

\paragraph{Example 6:  Image captioning and generation}

\begin{figure}[t]
    \centering
    \includegraphics[width=1\linewidth]{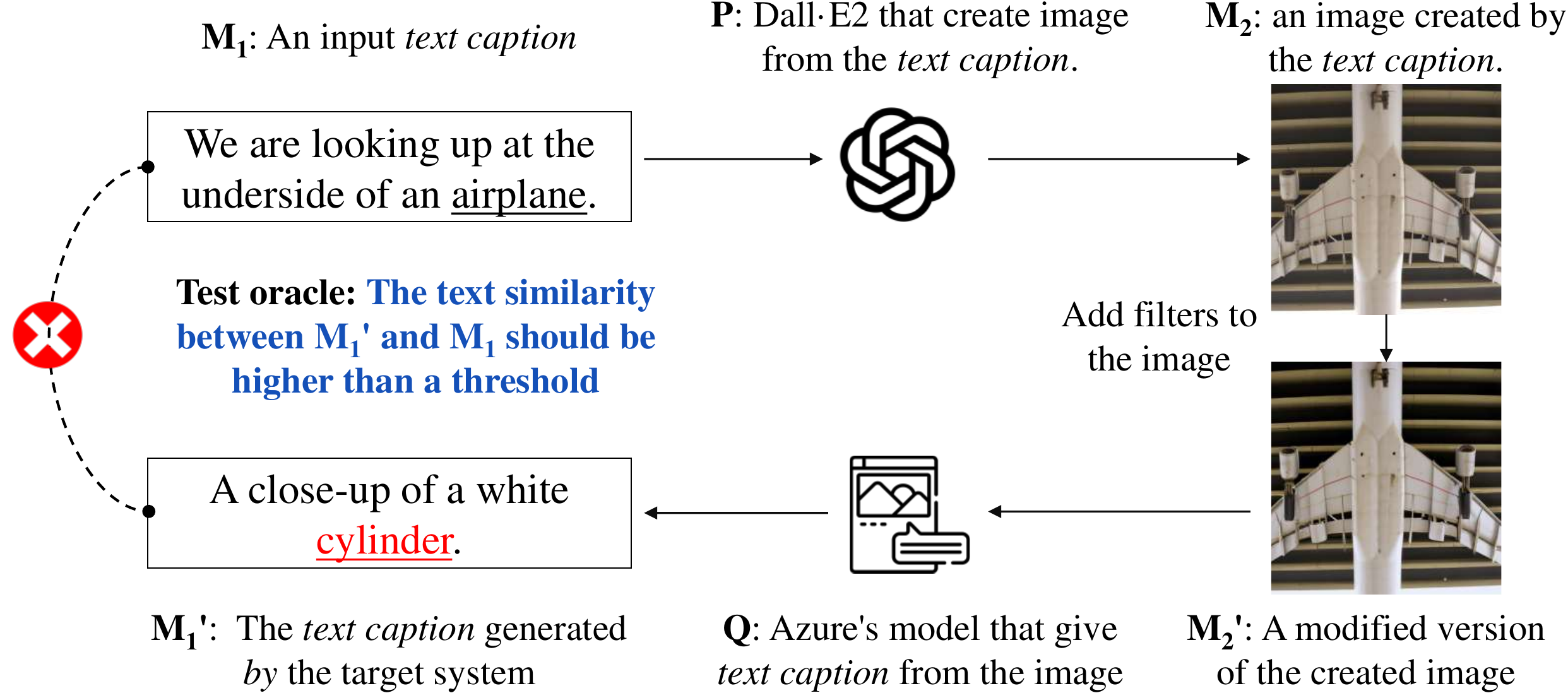}
    \caption{An illustrative example about using Retromorphic Testing to test Image generation and captioning.}
    \label{fig:text2image}
\end{figure}

Image captioning and generation play pivotal roles in computer vision and natural language processing.
Image captioning involves crafting textual depictions for images, while image generation entails crafting novel images based on textual descriptions.
We could adopt Retromorphic Testing for testing the image captioning system by using the image generation system as the auxiliary program, as shown in Fig~\ref{fig:text2image}.
We first generate a sentence $M_1$ as the input for the image generation system (DALL-E2\footnote{https://openai.com/dall-e-2/}) to generate an image $M_2$.
Then we reduce the brightness by $20\%$ and increase the contrast ratio by $10\%$  to obtain $M_2'$, which should not change the interpretation of the image.
After that, we use an image captioning system (\textit{i.e.,} Microsoft Azure Cognitive Services\footnote{https://azure.microsoft.com/en-gb/products/cognitive-services/vision-services/})  to generate a caption sentence $M_1'$ for $M_2'$.
We identify a Retromorphic Relation that the semantic of $M_1'$ should not deviate much from $M_1$, and use the text similarity to implement the test oracle, which should be higher than a threshold.
In practice, we use SimCSE~\cite{gao2021simcse} for calculating the similarity between the sentences, and set the threshold as 0.6.
In this case, $SimCSE(M_1,M_1')$ is 0.129 and lower than the threshold, which violates the Retromorphic Relation.
After manual inspection, we could see that the ``airplane'' within image $M_{2}$ is recognized as ``cylinder'' in $M_{1}'$.
This mismatch reflects an image captioning error.


\section{Challenges and Future Directions}

Despite the notable advantages of Retromorphic Testing, certain challenges need to be addressed.
First, the effectiveness of the test oracle highly depends on the auxiliary program, whose design demands a certain level of effort and favors developers who have a comprehensive understanding of the underlying systems.
Second, the creation of effective Retromorphic Relations is challenging. 
As depicted in Fig.~\ref{fig:threeTest} (c), Retromorphic Testing involves the transition between two modalities, allowing for data modification in the second modality. 
The effective design of test oracles requires a comprehensive consideration of both the modality transition and the alterations occurring in the second modality.


Future research can explore auxiliary program selection and construction, refining their ability to accurately reverse software outputs to original input modalities.
With the advancement of AI techniques such as Large Langauge Models, there is an increasing potential for uncovering a greater number of untapped Retromorphic Relations.

\section{Conclusion}


This paper introduces Retromorphic Testing, a new black-box testing methodology for tackling the oracle problem.
The core insight is to leverage an auxiliary program, form a dual-program structure in one of the three modes (\textit{i.e.,}  forward, backward, integrated), use the (optionally transformed) output of the forward program as the input of the backward program, and examine the retromorphic relation between the initial input and the output of the backward program in the input domain.
The testing scopes of Retromorphic Testing have been discussed in six testing instances ranging from integer factorization algorithm, database management system, to machine translators, demonstrating the wide applicability of the proposed concept. 
We believe Retromorphic Testing is a comparable methodology to differential testing and metamorphic testing and they complement with each other. 

\section{Data Availability}
Codes for this paper are available at an anonymous link.\footnote{https://github.com/CUHK-Shenzhen-SE/RetromorphicTesting} 






\balance
\bibliographystyle{ACM-Reference-Format}
\bibliography{software}

\end{document}